\documentclass[
  ,draft            
  ]
  {aipproc}

\layoutstyle{6x9}

\begin{document}

\title{KADoNiS- The Karlsruhe Astrophysical Database of Nucleosynthesis in Stars}


\author{I. Dillmann}{
  address={Institut f\"ur Kernphysik, Forschungszentrum Karlsruhe, Postfach 3640, D-76021 Karlsruhe,
  Germany}
  ,altaddress={Departement Physik und Astronomie, Universit\"at Basel, Klingelbergstrasse 82,
CH-4056 Basel, Switzerland} 
}
\author{M. Heil}{
  address={Institut f\"ur Kernphysik, Forschungszentrum Karlsruhe,
Postfach 3640, D-76021 Karlsruhe, Germany }}
\author{F. K\"appeler}{
  address={Institut f\"ur Kernphysik, Forschungszentrum Karlsruhe,
Postfach 3640, D-76021 Karlsruhe, Germany }}
\author{R. Plag}{
  address={Institut f\"ur Kernphysik, Forschungszentrum Karlsruhe,
Postfach 3640, D-76021 Karlsruhe, Germany }}

\author{T. Rauscher}{
  address={Departement Physik und Astronomie, Universit\"at Basel, Klingelbergstrasse 82,
CH-4056 Basel, Switzerland}}
\author{F.-K. Thielemann}{
  address={Departement Physik und Astronomie, Universit\"at Basel, Klingelbergstrasse 82,
CH-4056 Basel, Switzerland}}

\begin{abstract}
The "Karlsruhe Astrophysical Database of Nucleosynthesis in Stars"
(KADoNiS) project is an online database for experimental cross
sections relevant to the \emph{s} process and \emph{p} process. It
is available under
\url{http://nuclear-astrophysics.fzk.de/kadonis} and consists of
two parts. Part 1 is an updated sequel to the previous Bao et al.
compilations from 1987 and 2000 for (\emph{n},$\gamma$) cross
sections relevant to the big bang and \emph{s}-process
nucleosynthesis. The second part will be an experimental
\emph{p}-process database, which is expected to be launched in
winter 2005/06. The KADoNiS project started in April 2005, and a
first partial update is online since August 2005. In this paper we
present a short overview of the first update of the
\emph{s}-process database, as well as an overview of the status of
stellar (\emph{n},$\gamma$) cross sections of all 32 \emph{p}
isotopes.
\end{abstract}

\maketitle

\section{Stellar neutron capture compilations}
The first collection of stellar neutron capture cross sections was
published in 1971 by Allen and co-workers \cite{alle71}. This
paper reviewed the role of neutron capture reactions in the
nucleosynthesis of heavy elements and presented also of a list of
recommended (experimental or semi-empirical) Maxwellian averaged
cross sections at \emph{kT}= 30 keV (MACS30) for nuclei between
carbon and plutonium.

The idea of an experimental and theoretical stellar neutron cross
section database was picked up again by Bao and K\"appeler
\cite{bao87} for \emph{s}-process studies. This compilation
published in 1987 included cross sections for (\emph{n},$\gamma$)
reactions (between $^{12}$C and $^{209}$Bi), some (\emph{n,p}) and
(\emph{n},$\alpha$) reactions (for $^{33}$Se to $^{59}$Ni), and
also (\emph{n},$\gamma$) and (\emph{n},\emph{f}) reactions for
long-lived actinides. A follow-up compilation was published by
Beer, Voss and Winters in 1992 \cite{BVW92}.

In the update of 2000 this compilation \cite{bao00} was extended
to big bang nucleosynthesis. It now included a collection of
recommended MACS30 for isotopes between $^{1}$H and $^{209}$Bi,
and -- like the original Allen paper -- also semi-empirical
re\-commended values for nuclides without experimental cross
section information. These estimated values are normalized cross
sections derived with the Hauser-Feshbach code NON-SMOKER
\cite{rau95}, which account for known systematic deficiencies in
the nuclear input of the calculation. Additionally, the database
provided stellar enhancement factors and energy-dependent MACS for
energies between \emph{kT}= 5 keV and 100 keV.

The most recent KADoNiS version of this compilation has the aim to
provide a clearly arranged and user-friendly online database,
which is regularly updated and will be in later stages also
extended to \emph{p}-process studies.

\section{Part 1: Updated big bang and s-process database}
Included in the present update (status August 2005) were only
cross sections, which are already published. Six semi-empirical
estimates (see Table \ref{tab:a}) were replaced by experimental
data, and 20 recommended cross sections were updated by inclusion
of new measurements (Table \ref{tab:b}). A full list of
measurements with references, which were (will be) included in the
update(s) can be found on the KADoNiS homepage in the menu section
"Logbook".

Future efforts will be focussed on the re-evaluation of
semi-empirical cross sections, as well as the inclusion of
theoretical results derived with the Hauser-Feshbach code MOST
\cite{most05}. Another topic will be the re-calculation of cross
sections for isotopes, where a recent change in physical
properties (e.g. t$_{1/2}$, I$_\gamma$...) leads to changes in
already measured cross sections.

\begin{table}[!htb]
\begin{tabular}{cr@{$\pm$}lr@{$\pm$}l}
\hline \tablehead{1}{c}{b}{Isotope \\}
  & \tablehead{2}{c}{b}{Old recomm. value \\ $[mb]$}
  & \tablehead{2}{c}{b}{New exp. value \\ $[mb]$} \\
\hline
$^{128}$Xe & 248 ~&~ 66 & 262.5 ~&~ 3.7 \\
$^{129}$Xe & 472 ~&~ 71 & 617 ~&~ 12 \\
$^{130}$Xe & 141 ~&~ 51 & 132.0 ~&~ 2.1 \\
$^{147}$Pm & 1290 ~&~ 470 & 709 ~&~ 100 \\
$^{151}$Sm & 2710 ~&~ 420 & 3031 ~&~ 68 \\
$^{180}$Ta$^m$ & 1640 ~&~ 260 & 1465 ~&~ 100 \\
\hline
\end{tabular}
\caption{List of recommended semi-empirical stellar cross
sections, which were now replaced by experimental values.}
\label{tab:a}
\end{table}

\begin{table}[!htb]
\begin{tabular}{cr@{$\pm$}lr@{$\pm$}l}
\hline \tablehead{1}{c}{b}{Isotope \\ }
  & \tablehead{2}{c}{b}{Old recomm. value \\ $[mb]$}
  & \tablehead{2}{c}{b}{New recomm. value \\ $[mb]$} \\
\hline
$^{22}$Ne & 0.059~ & ~0.006 & 0.058~ & ~0.004 \\
$^{40}$Ar & 2.6~ & ~0.2 & 2.6~ & ~0.2 \\
$^{96}$Ru & 238~ & ~60 & 207~ & ~8 \\
$^{102}$Ru & 186~ & ~11 & 151~ & ~7 \\
$^{104}$Ru & 161~ & ~10 & 156~ &~5 \\
$^{110}$Cd & 246~ & ~10 & 237~ &~2 \\
$^{111}$Cd  & 1063~ & ~125 & 754~ & ~12 \\
$^{112}$Cd  & 235~ & ~30 & 187.9~ & ~1.7 \\
$^{113}$Cd  & 728~ & ~80 & 667~ & ~11 \\
$^{114}$Cd  & 127~ & ~5 & 129.2~ & ~1.3 \\
$^{116}$Cd & 59~ & ~2 & 74.8~ & ~0.9 \\
$^{135}$Cs & 198~ & ~17 & 160~ & ~10 \\
$^{139}$La & 38.4~ & ~2.7 & 31.6~ & ~0.8 \\
$^{175}$Lu & 1146~ & ~44 & 1219~ & ~10 \\
$^{176}$Lu & 1532~ & ~69 & 1639~ & ~14 \\
$^{176}$Hf & 455~ & ~20 & 626~ & ~11 \\
$^{177}$Hf & 1500~ & ~100 & 1544~ & ~12 \\
$^{178}$Hf & 314~ & ~10 & 319~ & ~3 \\
$^{179}$Hf & 956~ & ~50 & 922~ & ~8 \\
$^{180}$Hf & 179~ & ~5 & 157 ~& ~2 \\
\hline
\end{tabular}
\caption{List of previous and new recommended stellar cross
sections, which were updated by inclusion of new experimental
values.} \label{tab:b}
\end{table}

The KADoNiS homepage provides a datasheet with all necessary
informations for each isotope similar to the layout in
Ref.~\cite{bao00}. On the top of this page the recommended MACS30
for the total and all partial cross sections are shown. In the
"Comment" line one can find the previous recommended values,
special comments, and the date of the last review. The field "List
of all available values" includes the original values as given in
the respective publications, renormalized values, year of
publication, type of value (theoretical, semi-empirical or
experimental), a short comment about the method (accelerator,
neutron and reference source), and the (linked) reference(s).

This section is followed by the tabulated MACS, reaction rates and
stellar enhancement factors for energies between \emph{kT}= 5 and
100 keV. A "click" on the field "Show/hide mass chain" gives a
graphical plot of all available recommended total MACS30 for the
isotopic mass chain of the respective element. The bottom part of
each datasheet shows a chart of nuclides, which can be zoomed by
selecting different sizes (S, M, L, or XL). By clicking on an
isotope in this chart, one can easily jump to the respective
datasheet.

\section{Part 2: Experimental p-process database}
The second part of KADoNiS will be an experimental
\emph{p}-process database and is expected to be launched in winter
2005/06. It will be a collection of the available experimental
reaction rates relevant for \emph{p}-process studies, e.g.
($\gamma$,\emph{n}), ($\gamma$,$\alpha$), ($\gamma$,\emph{p}),
(\emph{n,p}), (\emph{n},$\alpha$), (\emph{p},$\alpha$), and their
inverse rates.

The role of (\emph{n},$\gamma$) reactions in the \emph{p} process
was early recognized by Rayet et al. \cite{ray90}. The
(\emph{n},$\gamma$)$\leftrightarrow$($\gamma$,\emph{n})
competition hinders the photodisintegration flux towards lighter
nuclei. Additionally the decrease in temperature at later stages
of the \emph{p} process leads to a freeze-out (T$_9$$\simeq$ 0.3,
corresponding to \emph{kT}= 25 keV) via neutron captures and
mainly $\beta^+$ decays, resulting in the typical \emph{p}-process
abundance pattern with maxima at $^{92}$Mo (\emph{N}=50) and
$^{144}$Sm (\emph{N}=82). The influence of a variation of reaction
rates on the final \emph{p} abundances has been demonstrated
repeatedly \cite{rau05,rapp04}.

Thus, it is necessary for \emph{p}-process studies to know the
neutron capture rates for both, at freeze-out energies (\emph{kT}=
25 keV) and at the \emph{p}-process energies (\emph{kT}= 170-260
keV). Table \ref{tab:c} gives an overview of the status of neutron
capture cross sections of all 32 \emph{p} nuclei at \emph{kT}= 30
keV. The Bao et al. compilation from 2000 \cite{bao00} provided
measured cross sections for 20 isotopes, but 9 of them
($^{92,94}$Mo, $^{96}$Ru, $^{124,126}$Xe, $^{130}$Ba, $^{156}$Dy,
$^{180}$W, and $^{190}$Pt) with uncertainties $\geq$9\%. For the
remaining 12 \emph{p} isotopes ($^{74}$Se, $^{84}$Sr, $^{98}$Ru,
$^{102}$Pd, $^{120}$Te, $^{132}$Ba, $^{138}$La, $^{158}$Dy,
$^{168}$Yb, $^{174}$Hf, $^{184}$Os, and $^{196}$Hg) only
theoretical predictions were available.

The (preliminary) results of our extended measuring program of
stellar neutron capture cross sections for \emph{p} nuclei are
shown in Table \ref{tab:c}. All of our measurements were carried
out on natural samples at the Karlsruhe 3.7 MV Van de Graaff
accelerator using the activation technique \cite{rapp02,dill05b}.
Neutrons were produced via the $^{7}$Li(\emph{p,n})$^{7}$Be
reaction by bombarding 30 $\mu$m thick layers of metallic lithium
on a water-cooled copper backing with protons of E$_p$= 1912 keV.
The resulting quasi-stellar neutron spectrum approximates a
Maxwellian distribution for \emph{kT}= 25.0 $\pm$ 0.5 keV
\cite{raty88}. In all eight cases ($^{74}$Se, $^{84}$Sr,
$^{96}$Ru, $^{102}$Pd, $^{120}$Te, $^{130,132}$Ba, and $^{174}$Hf)
we are able to reproduce the previous recommended total cross
sections from \cite{bao00} within 20\%. Thus, only 6 \emph{p}
isotopes ($^{98}$Ru, $^{138}$La, $^{158}$Dy, $^{168}$Yb,
$^{184}$Os, and $^{196}$Hg) remain without any experimental
stellar neutron cross section. With exception of $^{98}$Ru and
$^{138}$La, all isotopes can be measured with the activation
technique.

However, for an inclusion into the planned \emph{p}-process
database, those MACS30 have to be theoretically extrapolated to
\emph{p}-process temperatures. Another step is then the
calculation of inverse reaction rates by detailed balance.

\begin{table}
\begin{tabular}{cccccc}
\hline \tablehead{1}{c}{b}{Isotope}
  & \tablehead{2}{c}{b}{Hauser-Feshbach predictions}
  & \tablehead{2}{c}{b}{Recommended MACS30}
  & \tablehead{1}{c}{b}{Comments/} \\
  & \tablehead{1}{c}{b}{MOST \cite{most05}} & \tablehead{1}{c}{b}{NON-SMOKER \cite{rau01}} &
  \tablehead{1}{c}{b}{previous \cite{bao00}} & \tablehead{1}{c}{b}{new} & \tablehead{1}{c}{b}{Refs.} \\
  & \tablehead{1}{c}{b}{[mb]} & \tablehead{1}{c}{b}{[mb]} &
  \tablehead{1}{c}{b}{[mb]} & \tablehead{1}{c}{b}{[mb]} & \\
\hline
 $^{74}$Se & 247 & 207 & \emph{267 $\pm$ 25}\tablenote{Semi-empirical estimate.} & 276 $\pm$ 15 & \cite{dill05b}\\
 $^{78}$Kr & 388 & 351 & \multicolumn{2}{c}{312 $\pm$ 26} & \\
 $^{78}$Kr$\rightarrow$$^{m}$ & - & - & \multicolumn{2}{c}{92.3 $\pm$ 6.2} & \\
 $^{84}$Sr & 246 & 393 & \emph{368 $\pm$ 125}$^*$ & 302 $\pm$ 17 &  \cite{dill05b}\\
 $^{84}$Sr$\rightarrow$$^{m}$ & - & - & - & 190 $\pm$ 10 &  \cite{dill05b}\\
 $^{92}$Mo & 46 & 128 & \multicolumn{2}{c}{70 $\pm$ 10} & \\
 $^{94}$Mo & 85 & 151 & \multicolumn{2}{c}{102 $\pm$ 20} & \\
 $^{96}$Ru & 338 & 281 & 238 $\pm$ 60 & 207 $\pm$ 8 & \cite{rapp02}\\
 $^{98}$Ru & 358 & 262 & \multicolumn{2}{c}{\emph{173 $\pm$ 36}$^*$}& \\
 $^{102}$Pd & 670 & 374 & \emph{373 $\pm$ 118}$^*$ & 379 $\pm$ 16 & preliminary \\
 $^{106}$Cd & 365 & 451 & \multicolumn{2}{c}{302 $\pm$ 24} & \\
 $^{108}$Cd & 206 & 373 & \multicolumn{2}{c}{202 $\pm$ 9} & \\
 $^{113}$In & 316 & 1202 & \multicolumn{2}{c}{787 $\pm$ 70} & \\
 $^{113}$In$\rightarrow$$^{m}$ & - & - & \multicolumn{2}{c}{480 $\pm$ 160} & \\
 $^{112}$Sn & 154 & 381 & \multicolumn{2}{c}{210 $\pm$ 12} & \\
 $^{114}$Sn & 74 & 270 & \multicolumn{2}{c}{134.4 $\pm$ 1.8} & \\
 $^{115}$Sn & 247 & 528 & \multicolumn{2}{c}{342.4 $\pm$ 8.7} & \\
 $^{120}$Te & 309 & 551 & \emph{420 $\pm$ 103}$^*$ & 451 $\pm$ 18 & \cite{dill05a} prelim. \\
 $^{120}$Te$\rightarrow$$^{m}$ & - & - & - & 61 $\pm$ 2 & \cite{dill05a} prelim.\\
 $^{124}$Xe & 503 & 799 & \multicolumn{2}{c}{644 $\pm$ 83} & \\
 $^{124}$Xe$\rightarrow$$^{m}$ & - & - & \multicolumn{2}{c}{131 $\pm$ 17} & \\
 $^{126}$Xe & 335 & 534 & \multicolumn{2}{c}{359 $\pm$ 51} & \\
 $^{126}$Xe$\rightarrow$$^{m}$ & - & - & \multicolumn{2}{c}{40$\pm$6} & \\
 $^{130}$Ba & 493 & 730 & 760 $\pm$ 110 & 694 $\pm$ 20 & \cite{dill05a} prelim. \\
 $^{132}$Ba & 228 & 467 & \emph{379 $\pm$ 137}$^*$ & 368 $\pm$ 25 & \cite{dill05a} prelim. \\
 $^{132}$Ba$\rightarrow$$^{m}$ & - & - & - & 33.6 $\pm$ 1.7 & \cite{dill05a} prelim. \\
 $^{136}$Ce & 208 & 495 & \multicolumn{2}{c}{328 $\pm$ 21} & \\
 $^{136}$Ce$\rightarrow$$^{m}$ & - & - & \multicolumn{2}{c}{28.2 $\pm$ 1.6} & \\
 $^{138}$Ce & 61 & 290 & \multicolumn{2}{c}{179 $\pm$ 5} & \\
 $^{138}$La & 337 & 767 & \multicolumn{2}{c}{- \tablenote{No recommended value available, since $^{138}$La is of pure \emph{p}-process origin.}} & \\
 $^{144}$Sm & 39 & 209 & \multicolumn{2}{c}{92 $\pm$ 6} & \\
 $^{156}$Dy & 2138 & 1190 & \multicolumn{2}{c}{1567 $\pm$ 145} &  \\
 $^{158}$Dy & 1334 & 949 & \multicolumn{2}{c}{\emph{1060 $\pm$ 400}$^*$} & \\
 $^{162}$Er & 1620 & 1042 & \multicolumn{2}{c}{1624 $\pm$ 124} & \\
 $^{168}$Yb & 875 & 886 & \multicolumn{2}{c}{\emph{1160 $\pm$ 400}$^*$} & \\
 $^{174}$Hf & 763 & 786 & \emph{956 $\pm$ 283}$^*$ & 1056 $\pm$ 53 & preliminary \\
 $^{180}$W  & 751 & 707 & \multicolumn{2}{c}{536 $\pm$ 60} & \\
 $^{184}$Os & 709 & 789 & \multicolumn{2}{c}{\emph{657 $\pm$ 202}$^*$} & \\
 $^{190}$Pt & 634 & 760 & \multicolumn{2}{c}{677 $\pm$ 82} & \\
 $^{196}$Hg & 469 & 372 & \multicolumn{2}{c}{\emph{650 $\pm$ 82}$^*$} & \\
\hline
\end{tabular}
\caption{Status of MACS30 of all 32 \emph{p} nuclei. Recommended
cross section were taken from Ref. \cite{bao00}, unless another
reference is given.} \label{tab:c}
\end{table}

\begin{theacknowledgments}
We thank E. P. Knaetsch, D. Roller and W. Seith for their help
during the irradiations at the Van de Graaff accelerator. This
work was supported by the Swiss National Science Foundation Grants
2024-067428.01 and 2000-105328.
\end{theacknowledgments}


\bibliographystyle{aipproc}   


\end{document}